\documentclass[12pt]{article}
\usepackage{epsfig}
\usepackage{hyperref}
\def\beq{\begin{equation}}
\def\eeq#1{\label{#1}\end{equation}}
\def\eeqn{\end{equation}}

\def\beqa{\begin{eqnarray}}
\def\eeqa#1{\label{#1}\end{eqnarray}}
\def\eeqan{\end{eqnarray}}

\let\bar=\overbar

\def\Dslash{\not{\hbox{\kern-4pt $D$}}}
\def\dslash{\not{\hbox{\kern-2pt $\del$}}}

\def\msb{{\bar{\ssstyle M \kern -1pt S}}}

\def\neu#1{\widetilde\chi^0_{#1}}

\def\Title#1{\begin{center} {\Large {\bf #1} } \end{center}}
\def\Dzero{D/{\hbox{\kern-7pt O}}}
\def\ptmiss{/{\hbox{\kern-6.5pt $\rm{p_T}$}}}
\def\chpm#1{\widetilde\chi^{\pm}_{#1}}
\newcommand{\hlink}[1]{\href{#1}{#1}}

\begin{document}

\hspace {3.5in} FERMILAB-CONF-08-577-E

\Title{Beyond the Standard Model - Searches at HERA and the Tevatron}

\bigskip\bigskip

%+\addtocontents{toc}{{\it D. Reggiano}}
%+\label{ReggianoStart}

\begin{raggedright}  

{\it Stefan Gr\"unendahl\index{Gr\"unendahl, S.}\\
Fermi National Accelerator Laboratory\\
PO Box 500, Batavia, IL 60510, USA}
\bigskip\bigskip
\end{raggedright}

\section{Introduction}
Searches for Physics beyond the Standard Model have entered an exciting new phase:  the complete HERA \index{HERA} data samples 
 obtained until the end of operations in the Summer of 2007 are now available for analysis.  ZEUS \index{ZEUS} and H1 \index{H1} have each collected about  $0.5\,fb^{-1}$ of lepton proton data, distributed over electron and positron running, and over different lepton beam polarisations (see Table~\ref{tab:Hera_lum}).
At the same time the Tevatron \index{Tevatron} proton-antiproton collider is accumulating data at unprecedented rates, with current analyses based on up to $3\,fb^{-1}$. 
The Tevatron experiments \Dzero\  \index{DZERO} and CDF \index{CDF} have each already recorded over $4\,fb^{-1}$  (Fall 2008), and are aiming for a total of $8\,fb^{-1}$ of $\bar{p} p$ collisions at 2 TeV center-of-mass energy for Tevatron Run II. \index{Tevatron Run II}

I am presenting recent updates (from the last 12 months) on searches, grouped loosely into three classes: well-established `traditional' searches, mostly for very specific signatures and models, more recent and/or more generalized searches for broader classes of phenomena, and newer searches  that strive to be model-independent.
Table~\ref{tab:summary} provides an overview of the analysis updates presented here.

\begin{table}[htbp]
   \begin{center}
      \begin{tabular}{ |c|c|c|c|c|c|r|} 
   \hline
      exp. & analysis & channel &  $\mathcal{L}$ [$fb^{-1}$] & updated & limit/result   & ref. \\
         \hline
%6 slide number
      \Dzero\ &  LQ1& 2 e + 2 jets  & 1     & 6/14/08 & $m_{LQ1}>292\,GeV$ & \cite{dzero:LQ1} \\  %\hline
        \hline
      %7
      H1 &  LQ1& e + jet  & 0.449  & 4/08  & $m_{LQ1}\ \rm{vs}\  \lambda$ (contour) & \cite{H1:LQ1} \\ %  \hline
         \hline
      %8
      &&$2 \mu $+ 2 jets &&&&\\
            \raisebox{1.5ex}[0pt]{\Dzero\ } &  \raisebox{1.5ex}[0pt]{ LQ2}&  $ \mu \nu $ + 2 jets  & \raisebox{1.5ex}[0pt]{1 }   & 
            \raisebox{1.5ex}[0pt]{8/29/08} &  \raisebox{1.5ex}[0pt]{$m_{LQ2}>316\,GeV$} &  \raisebox{1.5ex}[0pt]{\cite{dzero:LQ2}}\\  % \hline
           \hline
      %9
         &&&&& $m_{LQ1/2}> 179\,GeV$ &\\
      \raisebox{1.5ex}[0pt]{CDF }&  \raisebox{1.5ex}[0pt]{ LQ }& \raisebox{1.5ex}[0pt]{ 2 jets + \ptmiss} 
            &  \raisebox{1.5ex}[0pt]{2}     &  \raisebox{1.5ex}[0pt]{4/03/08} &  $m_{LQ3}> 169\,GeV$ 
             &  \raisebox{1.5ex}[0pt]{\cite{cdf:LQ}}\\  
               \hline
      %10
      \Dzero\ &  LQ3 & $\tau$ + b  & 1.05  & 6/21/08  & $m_{LQ3}> 210\,GeV$ & \cite{dzero:LQ3} \\   %\hline
        \hline
      %12
         && $\rho_T \rightarrow \pi_T W $ &&&&\\ 
       \raisebox{1.5ex}[0pt]{CDF} &  \raisebox{1.5ex}[0pt]{TC} & $ l\nu b\bar{b})$ &
        \raisebox{1.5ex}[0pt]{1.9}  & \raisebox{1.5ex}[0pt]{4/11/08}  & \raisebox{1.5ex}[0pt]{contour} & \raisebox{1.5ex}[0pt]{\cite{cdf:technirho}} \\  
          \hline
       %13
      CDF &  mSUGRA&$ \chpm1\neu2 $ & 2  & 1/10/08 & $m_{\chpm1}> 145\,GeV$ & \cite{cdf:trilepton} \\  
        \hline
      \Dzero &  mSUGRA&trileptons & $\le 1.7$  & 7/31/07  & $m_{\chpm1}> 145\,GeV$ & \cite{dzero:trilepton} \\  
        \hline
      %14
      &&&&&$m_{\widetilde{q}}=m_{\widetilde{g}}> 392\,GeV$& \\
      \raisebox{1.5ex}[0pt]{CDF} &  \raisebox{1.5ex}[0pt]{$\widetilde{q},\widetilde{g}$}&  \raisebox{1.5ex}[0pt]{jets + \ptmiss}
        & \raisebox{1.5ex}[0pt]{2}  & \raisebox{1.5ex}[0pt]{2/14/08}  & \parbox{90pt}{ $m_{\widetilde{g}}> 280\,GeV$ for $m_{\widetilde{q}} \le 600\,GeV$}
        & \raisebox{1.5ex}[0pt]{\cite{cdf:squark}} \\  
          \hline
       \Dzero & $\widetilde{q},\widetilde{g}$& jets + \ptmiss  & 2.1  & 1/24/08  &  \parbox{90pt}{ $m_{\widetilde{g}}> 308\,GeV$  
       $m_{\widetilde{q}}> 379\,GeV$} & \cite{dzero:squark} \\  
          \hline
       %15
     {CDF} & {stop}& $(e,\mu)+ b + \ptmiss $ & {2.7}  &{7/22/08}       & {contour} & {\cite{cdf:stop}} \\  
        \hline
      \Dzero &  stop& \parbox{50pt}{2 c + \ptmiss}  & 1  & 3/14/08  & contour & \cite{dzero:stop_c} \\  
        \hline
      \Dzero &  stop& \parbox{50pt}{e + $\mu$ + 2b + \ptmiss}  & 1.1  & 4/1/08  & contour & \cite{dzero:stop_emu2b} \\  
        \hline
      %16
      &&&&&$m_{\neu1}> 125\,GeV$&\\
      \raisebox{1.5ex}[0pt]{\Dzero} &  \raisebox{1.5ex}[0pt]{GMSB}& \raisebox{1.5ex}[0pt]{di-photon}  & \raisebox{1.5ex}[0pt]{1.1}  
      & \raisebox{1.5ex}[0pt]{10/21/07}  & $m_{\chpm1}> 229\,GeV$  & \raisebox{1.5ex}[0pt]{\cite{dzero:diphoton}} \\  
        \hline
      %17
      \Dzero &  $W^\prime$ &$e\nu$ & 1  & 10/16/07  & $m_{W^\prime}> 1 TeV$ & \cite{dzero:wprime} \\  
        \hline
      %18
      CDF &  $W^\prime$ & $t \bar{b}$  & 1.9  & 12/20/07  & $m_{W^\prime}> 800\,GeV$ & \cite{cdf:wprime} \\  
        \hline
      \Dzero &  $W^\prime$ & $t \bar{b}$ & 0.9  & 5/30/08  & $m_{W^\prime}> 731\,GeV$ & \cite{dzero:wptb} \\  
        \hline
      %19 
      CDF &  $Z^\prime$& 2 e  & 2.5  & 3/6/08  & $m_{Z^\prime}> 966\,GeV$ & \cite{cdf:zprime} \\ 
        \hline
      %20 
      CDF &  LED & mono-$\gamma$/jet  & 2/1.1  & 2/21/08  & $m_{D}> 950\,GeV$ (n=6) & \cite{cdf:LED} \\  
        \hline
      \Dzero &  LED & mono-$\gamma$ & 2.7  & 7/23/08  & $m_{D}> 831\,GeV$ (n=6) & \cite{dzero:LED} \\  
        \hline
      %21
      \Dzero &  Z+$\gamma$& $\gamma + ee$ or $\mu\mu$  & 1.1/1.0  & 6/3/08  & contour & \cite{dzero:zgamma} \\  
        \hline
      %22
      CDF &  $t^\prime \bar{t^\prime}$& $e\nu jjjj$  & 2.3  & 3/7/08  & $m_{t^\prime}> 284\,GeV$ & \cite{cdf:tprime} \\  
        \hline
      %23
      CDF & {same sign tt}& various  & 2  & 5/8/08  & \parbox{80pt}{$\xi < 0.85$ for $m_{\eta} = 200\,GeV$} & \cite{cdf:maxflav} \\  
        \hline
      %25
      CDF & $q^{*}$ & {dijets}  & 1.1  & 3/19/08  &{$260 < m_{q^{*}}< 870\,GeV$} & \cite{cdf:dijet} \\  
        \hline
      %26
      \Dzero &  $e^*$ &  $ee\gamma$ & 1  & 1/6/08  & $m_{e^*}> 756\,GeV$ & \cite{dzero:estar} \\  
        \hline
      %27
      H1 &  $e^*$& $e\gamma, eZ, \nu W$ & 0.475  & 5/08  & contour & \cite{H1:estar} \\  
        \hline
      %30
      H1 &  $\nu^*$&$ \nu \gamma, \nu Z, e W$  & 0.184  & 2/08  & contour & \cite{H1:nustar} \\  
        \hline
      %32
      Zeus & {iso.leptons}  & $e/\mu$+ \ptmiss  & 0.5  & 7/08  & no excess & \cite{Zeus:isolep} \\  
        \hline
      %33
      Hera &  {multileptons} & leptons & 0.5  & 6/08  & no excess & \cite{Hera:mullep} \\  
        \hline
      %35
      \Dzero &  \parbox{50pt}{di-em vertices} & 2e/2$\gamma$  & 1.1  & 6/13/08  & limits & \cite{dzero:loli} \\  
        \hline
      %36,%40
      CDF &  {Vista/Sleuth} & various  & 2.0  & 2/28/08  & no dev. found & \cite{cdf:vista} \\  
               \hline
   \end{tabular}
   \caption{Searches Summary as of Summer 2008}
   \label{tab:summary}
      \end{center}
\end{table}

\begin{table}[htbp]
\begin{center}
\begin{tabular}{|c|c|c|c|c|c|}
\hline
 &  & Electron energy & Proton energy & $\sqrt{s}$ & $\mathcal{L}$\\
 \raisebox{1.5ex}[0pt]{Period} & \raisebox{1.5ex}[0pt]{Beams}  &  (GeV) & (GeV) &  (GeV) & ($\mathrm{pb}^{-1}$) \\  
\hline
1994--1997 & $e^+p$ & $27.5$ & $820$ & $300$ & $48.2$ \\
1998--1999 & $e^-p$ & $27.5$ & $920$ & $318$ & $16.7$ \\
1999--2000 & $e^+p$ & $27.5$ & $920$ & $318$ & $65.1$ \\
\hline
2003--2004 & $e^+p$ & $27.5$ & $920$ & $318$ & $40.8$ \\
2004--2006 & $e^-p$ & $27.5$ & $920$ & $318$ & $190.9$ \\
2006--2007 & $e^+p$ & $27.5$ & $920$ & $318$ & $142.4$ \\

\hline
\end{tabular}
\caption{Zeus beam configurations time line, from \protect\cite{Zeus:isolep}. \index{HERA data samples}}
\label{tab:Hera_lum}
\end{center}
\end{table}

\section{Specialised Searches}
\subsection{Leptoquarks}
\begin{figure}[htb]
\begin{minipage} [l] {0.5\textwidth}
\begin{center}
\epsfig{file=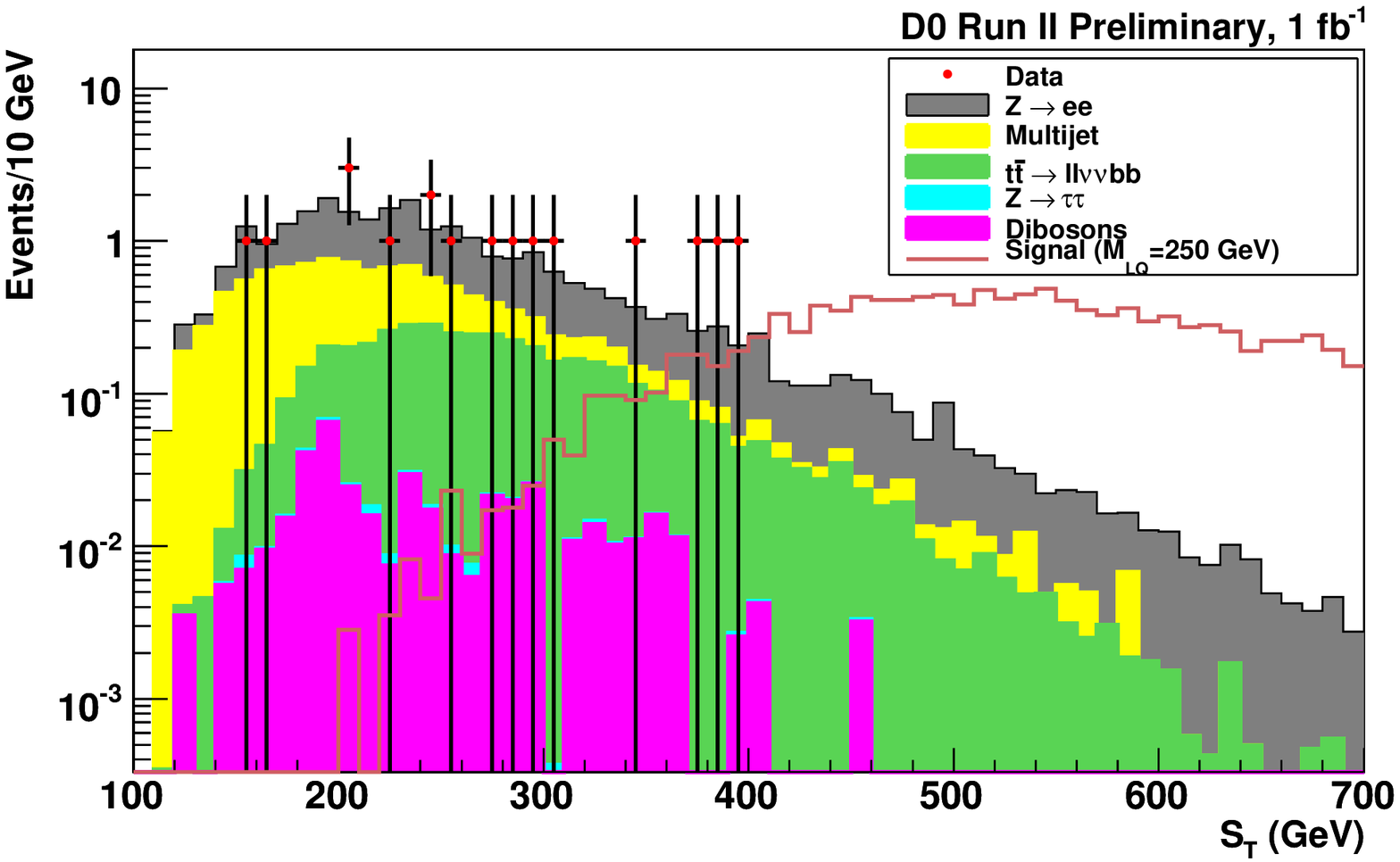,width=2.5in}
\caption{Sum of LQ decay transverse energies}
\label{fig:N62F2b}
\end{center}
\end{minipage}
%\end{figure}
\hspace{0.1cm}
%\begin{figure}[htb]
\begin{minipage} [r] {0.5\textwidth}
\begin{center}
\epsfig{file=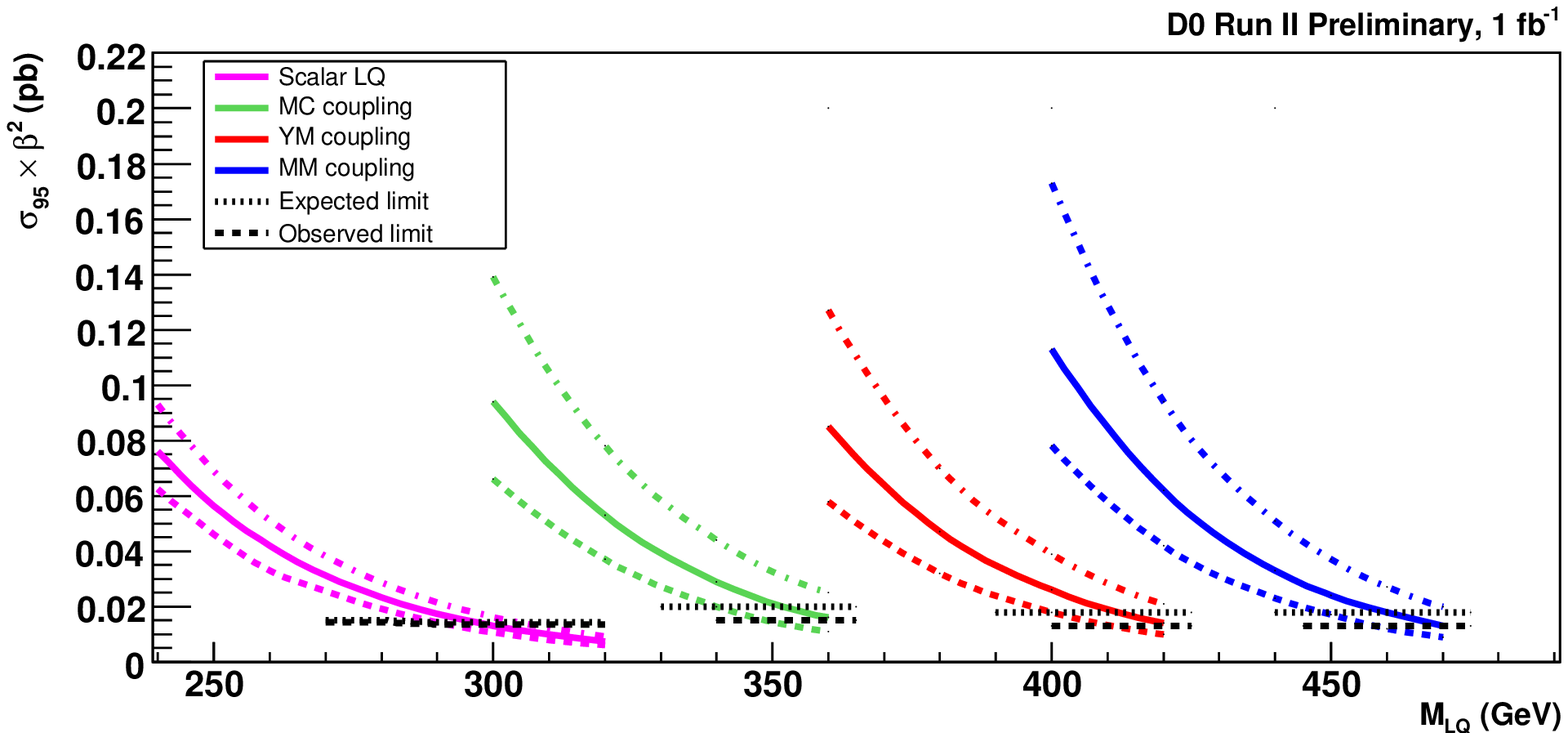,width=3in}
\caption{\Dzero\ LQ1 exclusion limits}
\label{fig:N62F6}
\end{center}
\end{minipage}
\end{figure}

\begin{figure}[htb]
\begin{minipage} [l] {0.5\textwidth}
\begin{center}
\epsfig{file=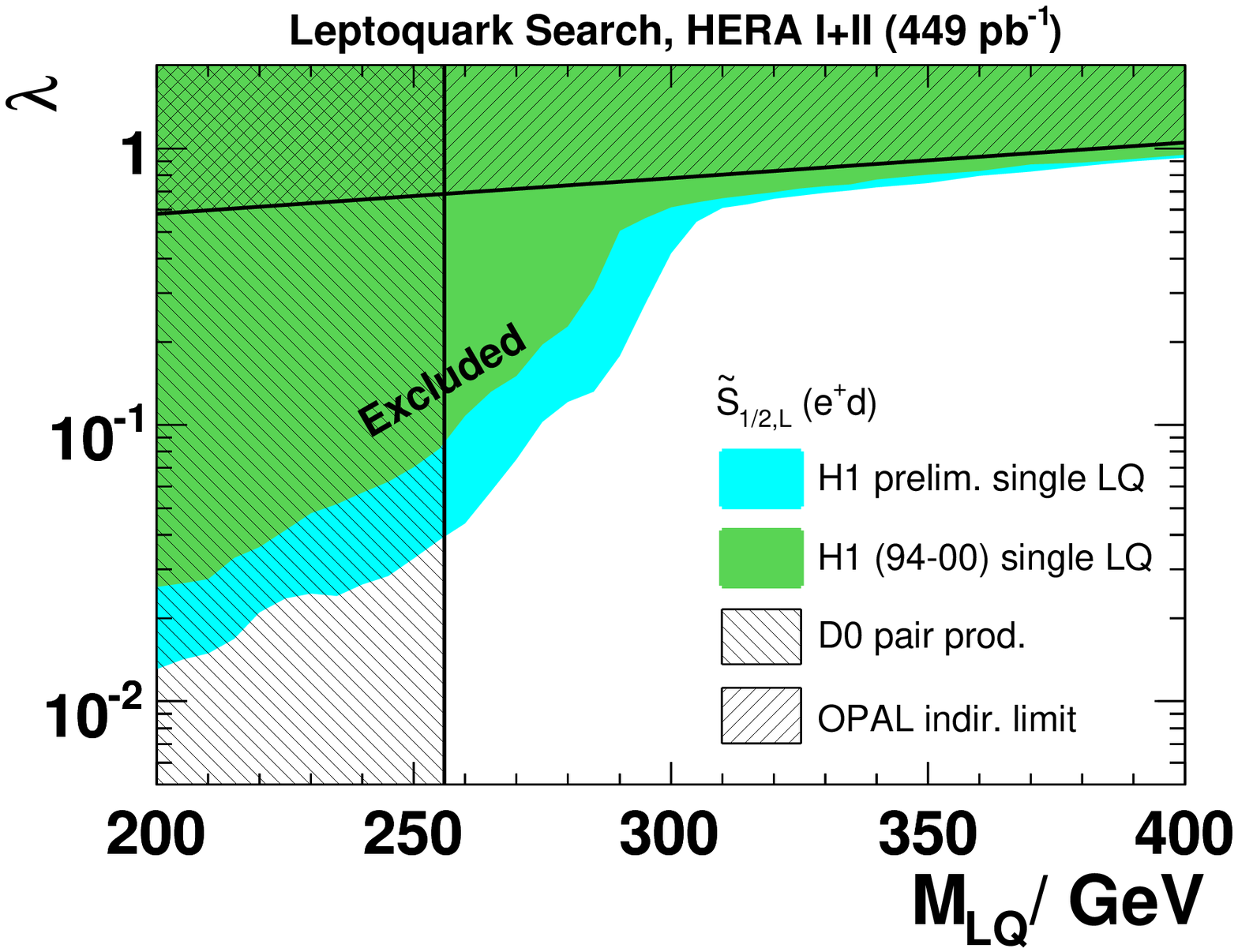,width=3in}
\caption{H1 LQ1 exclusion limits}
\label{fig:H1LQ1}
\end{center}
\end{minipage}
\hspace{0.1cm}
\begin{minipage} [r] {0.5\textwidth}
\begin{center}
\epsfig{file=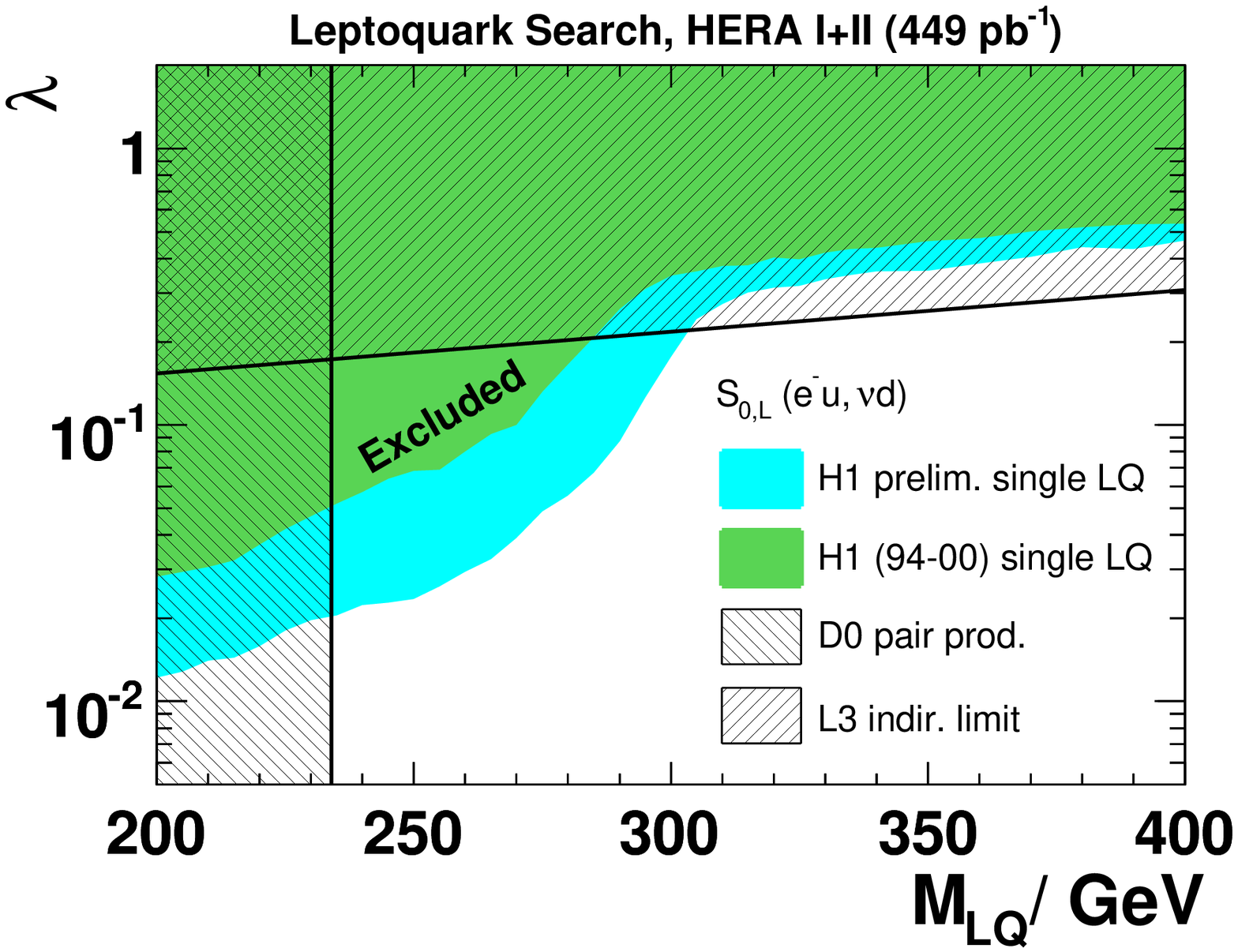,width=3in}
\caption{H1 LQ1 exclusion limits}
\label{fig:H1LQ1b}
\end{center}
\end{minipage}
\end{figure}

Leptoquarks \index{Leptoquarks} carry both lepton and quark quantum numbers and are present in many Grand Unified Theories, 
Technicolor and other extensions of the Standard Model. At a hadron collider, leptoquarks of all three generations are produced in pairs via the strong interaction.
At Hera production of the first generation LQ1 is proportional to the non-SM lepton-quark coupling. D0 has updated its search for leptoquarks of the first generation, coupling to electrons and up and down quarks, with results based on  $1\,fb^{-1}$ \cite{dzero:LQ1}. Fig.~\ref{fig:N62F2b} compares the distribution of the sum of the transverse energies of leptoquark decay products for SM backgrounds and for a hypothetical 250\,GeV leptoquark; no excess is visible. Scalar leptoquarks with masses less than 292\,GeV are excluded at 95\% confidence level (cf.~Fig.~\ref{fig:N62F6}). H1 \cite{H1:LQ1} has updated its LQ1 results a while ago. Fig.~\ref{fig:H1LQ1} shows the exclusion contours for two different LQ variants.

Both \Dzero\ and CDF have also updated their results for second generation leptoquark searches. While the \Dzero\ analysis \cite{dzero:LQ2}
requires a single muon accompanied by two jets and missing transverse energy \ptmiss, CDF searches in the (exclusive) dijet plus missing transverse energy channel and thus sets limits on first, second and third generation leptoquarks \cite{cdf:LQ}. For results see the references and table \ref{tab:summary}.

\begin{figure}[htb]
\begin{minipage} [l] {0.5\textwidth}
\begin{center}
\epsfig{file=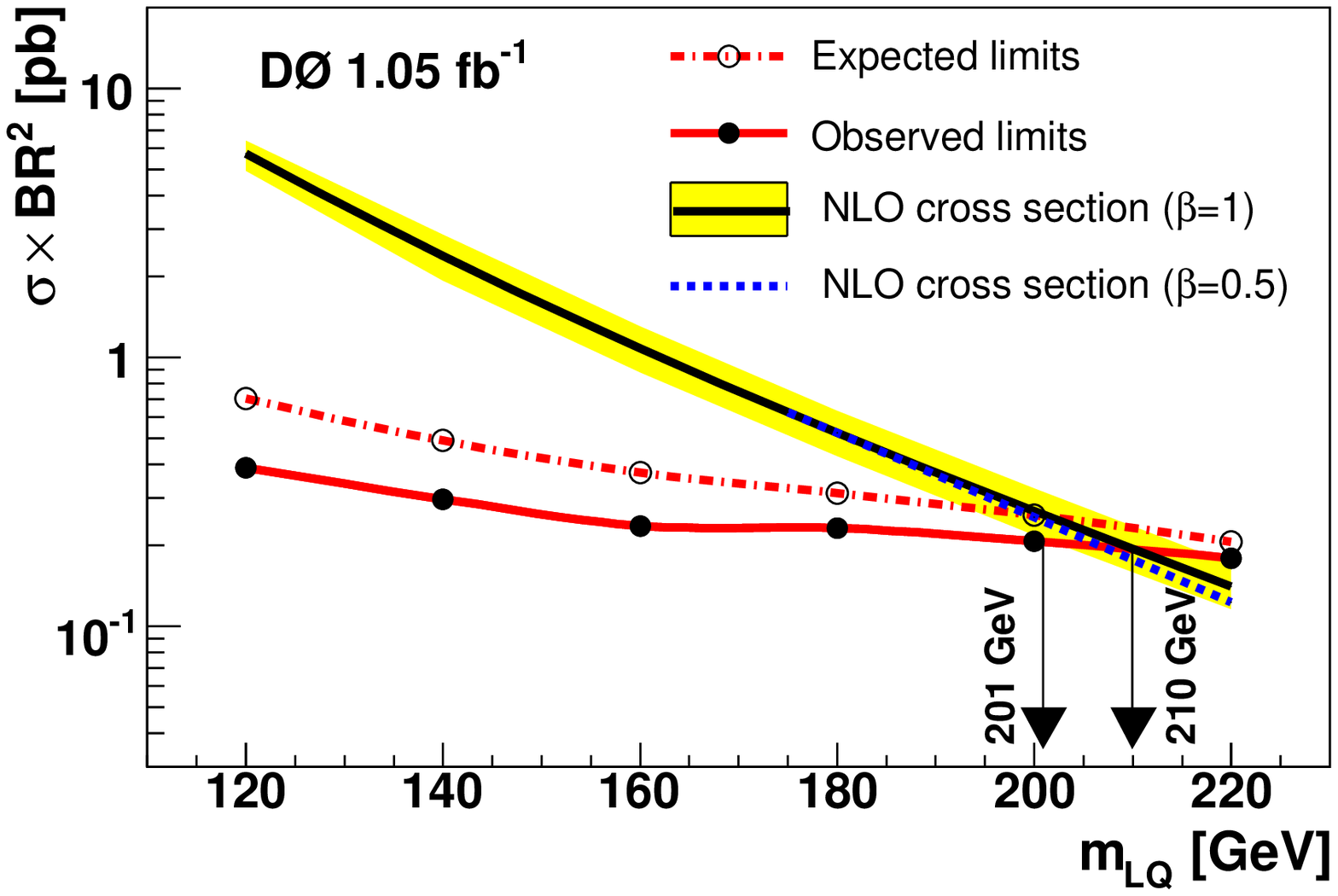,width=3.4in}
\caption{\Dzero\ LQ3 exclusion limit}
\label{fig:dzeroLQ3}
\end{center}
%\end{figure}
\end{minipage}
\hspace{0.1cm}
\begin{minipage} [r] {0.5\textwidth}
%\begin{figure}[htb]
\begin{center}
\epsfig{file=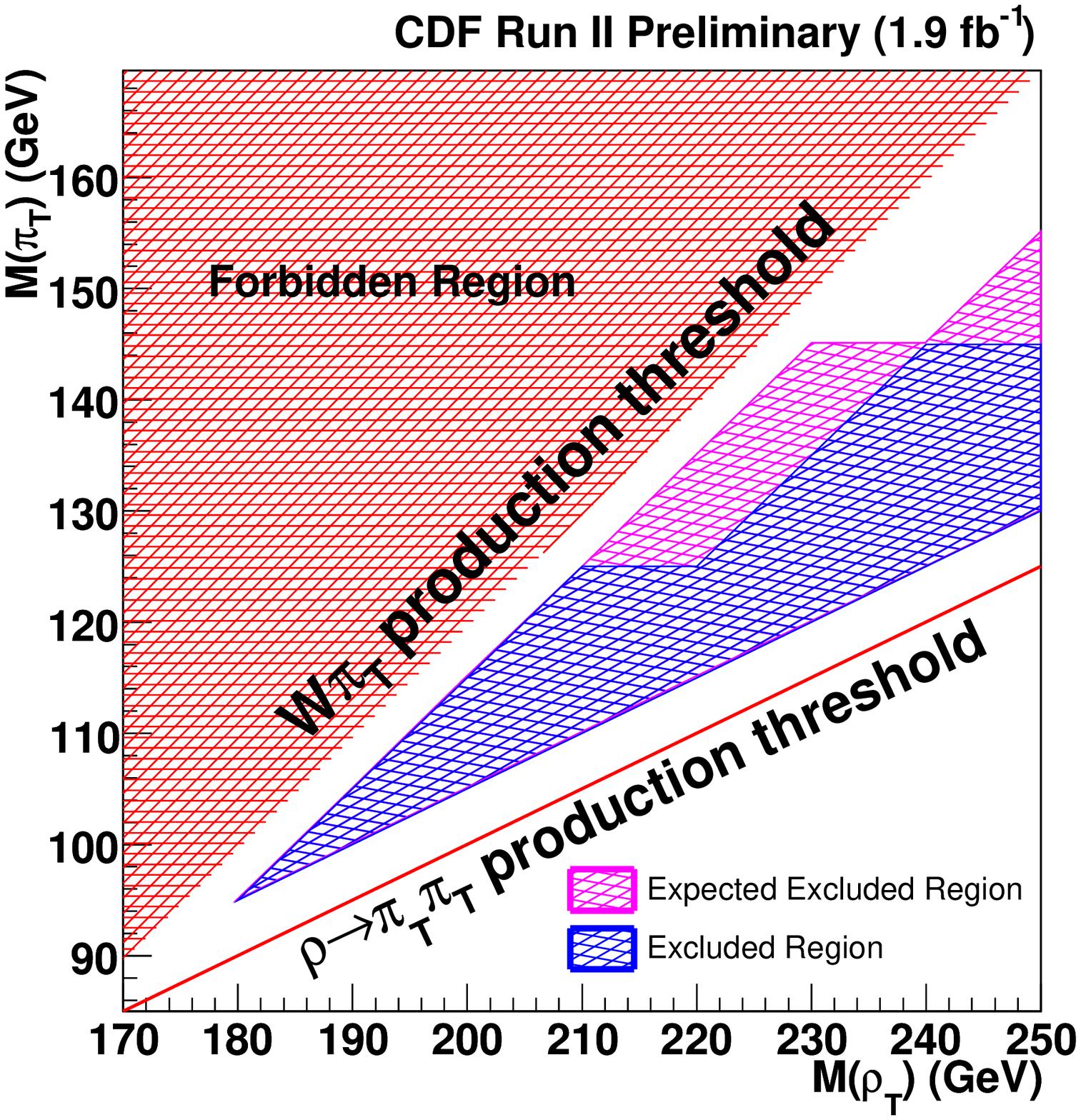,width=2.6in}
\caption{ CDF Techni-$\rho$ exclusion limit}
\label{fig:cdftechnirho}
\end{center}
\end{minipage}
\end{figure}

\Dzero\  has a recent $\tau$ based analysis for third generation leptoquarks \cite{dzero:LQ3}, which yields a limit of 210\,GeV for a third generation leptoquark at a branching fraction of $\beta=1$ into the charged lepton channel (Fig.~\ref{fig:dzeroLQ3}).

\subsection{Technicolor \index{Technicolor}}
Pursuing a signal very similar to that for associated production of W + Higgs, CDF has recently studied the decay of a hypothetical techni-$\rho$ 
\index{Technirho} into a techni-pion plus W and then further into W plus b jets \cite{cdf:technirho}. From data corresponding to $1.9\,fb^{-1}$, CDF excludes a new region in the $m_{\rho_T}$ - $m_{\pi_T}$ plane (Fig.~\ref{fig:cdftechnirho}).

\subsection{SUSY}
Ever since its invention in the last century, supersymmetry \index{Supersymmetry} \index{SUSY} has been the most  copious source of hypothetical particles. 
\subsubsection{Trileptons}
A stalwart among the searches beyond the standard model is the quest for chargino \index{Chargino} and neutralino \index{Neutralino} pair production via decay of an off-shell W. Both CDF and \Dzero\ have updates to their trilepton \index{Trilepton} analyses available \cite{cdf:trilepton,dzero:trilepton}. Masses up to about 145\,GeV are excluded.

\subsubsection{Squarks and Gluinos}
More stringent (higher) limits can be set on squarks \index{Squarks} and gluinos, \index{Gluinos} due to their strong production. With a little bit more than 2 inverse femtobarns analysed, \Dzero\ excludes squarks below 392\,GeV and gluinos below 327\,GeV \cite{dzero:squark}, and CDF \cite{cdf:squark} is not far behind.

\subsubsection{Stop \index{Stop}}
In some models, the supersymmetric partner of the top quark is the lightest of the squarks. Among them, CDF and \Dzero\ have studied three different potential decay paths of this hypothetical particle:
\begin{enumerate}
\item $\tilde{t}_1 \rightarrow c \neu1 \rightarrow $ 2 c jets $+ \ptmiss$ \cite{dzero:stop_c}
\item $\tilde{t}_1 \rightarrow b \chpm1 \rightarrow b \neu1 l \nu  \rightarrow $ 2 leptons (1 isolated), 2 jets (1 btag), $\ptmiss$ \cite{cdf:stop}
\item $\tilde{t}_1 \rightarrow b l \widetilde{\nu} \rightarrow e + \mu + 2 b + \ptmiss$ \cite{dzero:stop_emu2b}
\end{enumerate}
Regions of SUSY parameter space extending almost to $m_{\tilde{t}_1} = 180\, GeV$ have been excluded.

\subsubsection{Gauge-Mediated Symmetry Breaking}
\Dzero\ is making extended use of the `EM pointing' technique \index{EM pointing} \index{photon pointing} it pioneered in Run I, using the lateral and depth segmentation of its electromagnetic calorimeter to associate electromagnetic showers with a particular primary vertex (Fig.~\ref{fig:EM_pointing}). This helps in suppressing multiple interaction background, which can otherwise become a problem at the high luminosities (several $10^{32}s^{-1}cm^{-2}$) the Tevatron \index{Tevatron luminosity} is routinely achieving these days. 
Using this technique, a search for signs of GMSB \index{GMSB} yielding two photons plus missing transverse energy based on $1.1\,fb^{-1}$ sets limits of $m_{\chpm1}> 229\,GeV$  and $m_{\neu1}> 125\,GeV$ \cite{dzero:diphoton}.

\subsection{$\rm{W}^\prime$}
Only slightly younger than the W itself are searches for its heavier reincarnation, \index{reincarnation} the $\rm{W}^\prime$. \index{Wprime} The latest \Dzero\  result in the electron channel
\cite{dzero:wprime} pushes the mass limit into the TeV region. Both \Dzero\ and CDF are also searching for a  $\rm{W}^\prime$ decaying to top plus bottom \cite{cdf:wprime,dzero:wptb}. Here the limits are still a bit shy of the TeV boundary.

\subsection{$\rm{Z}^\prime$ \index{Zprime}}
The only update falling into our time window is from CDF:  in \cite{cdf:zprime} CDF searches the $e^+e^-$ mass spectrum for deviations from the standard model prediction, using narrow resonances (width compatible with the detector mass resolution) as templates. The largest deviation found has a significance of $2.5\,\sigma$. 
A $\rm{Z}^\prime$ with SM couplings is ruled out up to a mass of 966\,GeV.
  
\section{Recent Additions to the Searches Repertoire}
\subsection{Large Extra Dimensions}
Both \Dzero\ and CDF search for large extra dimensions \index{Large Extra Dimensions} \index{LED} in the monophoton and monojet final states. \Dzero\ recently updated their monophoton \index{monophotons} analysis (again using EM pointing) \cite{dzero:LED} based on a sample corresponding to $1.05\,fb^{-1}$, and CDF has new combined results for monophotons and monojets \index{monojets} using $1.1\,fb^{-1}$ and $2.0\,fb^{-1}$, respectively. Lower limits on the $M_D$ parameter are about 1 TeV, depending on the number of extra dimensions, and are better than the combined LEP limits for $4\le n_D \le 8$.
\subsection{$Z + \gamma$ \index{$Z + \gamma$}}
\begin{figure}[htb]
\begin{minipage} [l] {0.5\textwidth}
\begin{center}
\epsfig{file=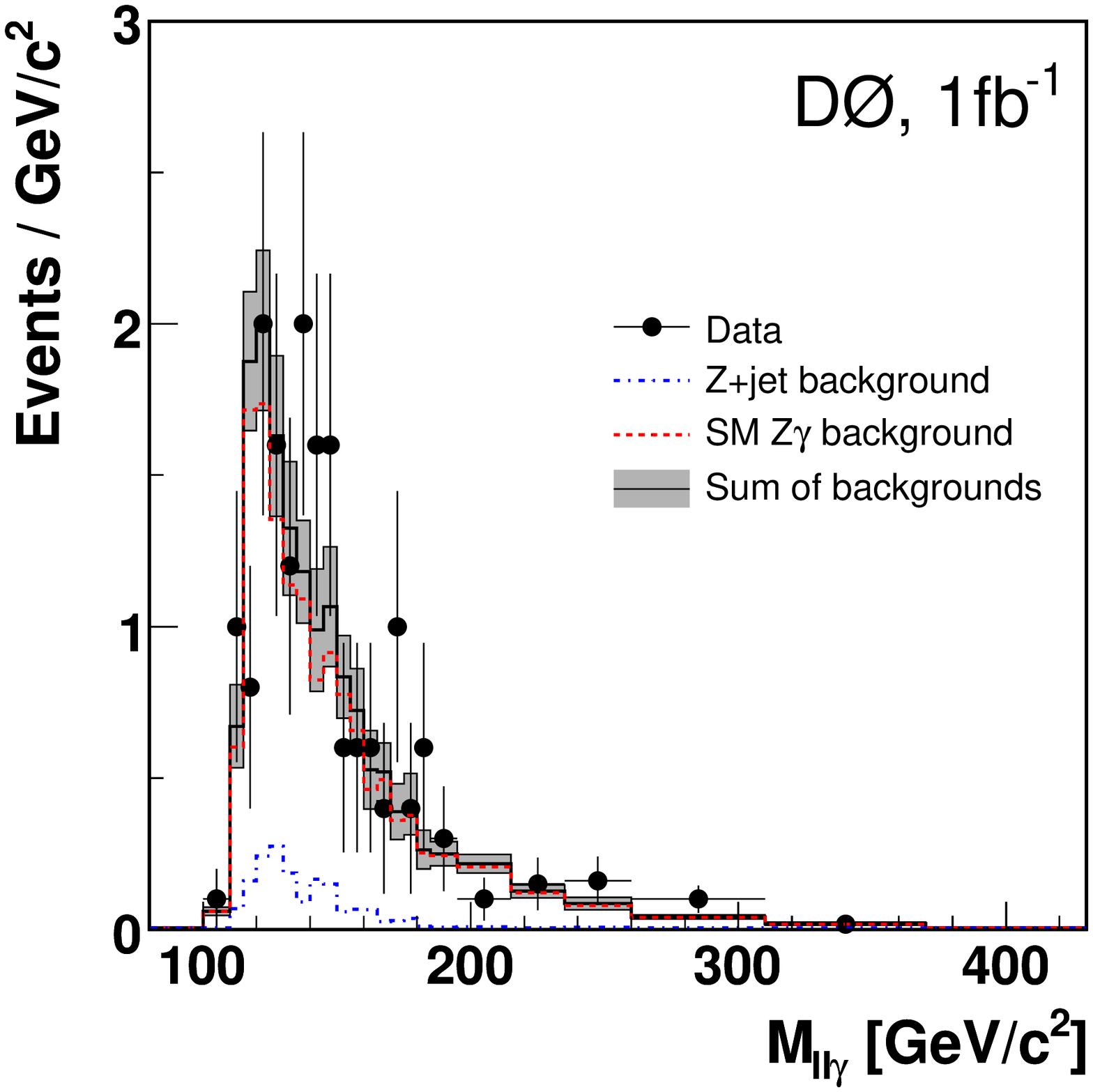,width=2.2in}
\caption{\Dzero\  $Z + \gamma$ mass spectrum for $ee\gamma$ and $\mu \mu \gamma$ combined}
\label{fig:dzero_zgamma}
\end{center}
\end{minipage}
\hspace{0.1cm}
\begin{minipage} [r] {0.5\textwidth}
\begin{center}
\epsfig{file=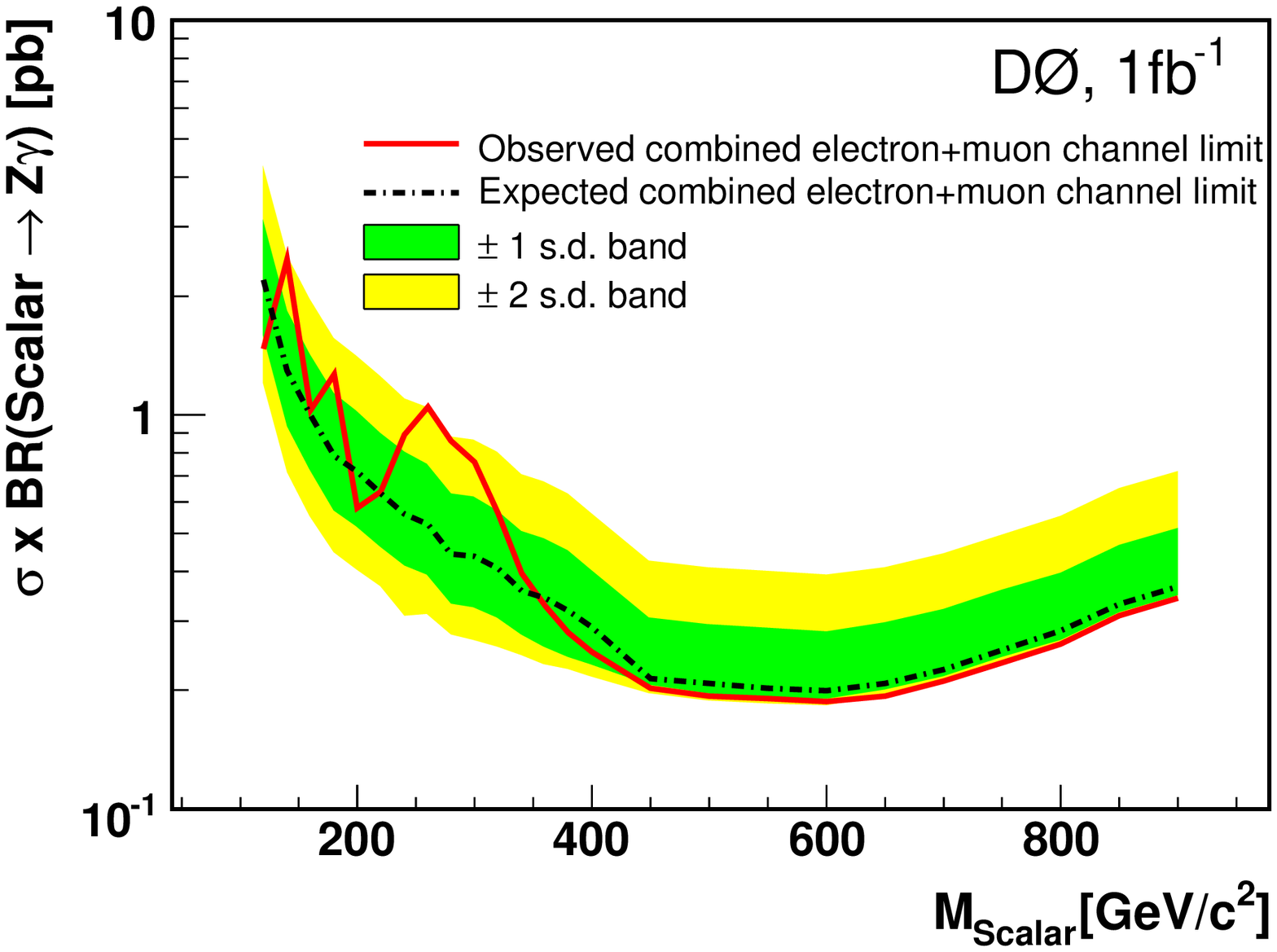,width=3in}
\caption{\Dzero\  $Z + \gamma$  exclusion contour}
\label{fig:dzero_zgamma_b}
\end{center}
\end{minipage}

\end{figure}

\Dzero\ has published a search for scalar or vector particles decaying into Z $+ \gamma$, based on data corresponding to $1\,fb^{-1}$ \cite{dzero:zgamma}. Fig.~\ref{fig:dzero_zgamma} shows the invariant mass spectrum and  Fig.~\ref{fig:dzero_zgamma_b} the resulting exclusion limit. 

\subsection{Fourth Generation}
\begin{figure}[htb]
\begin{minipage} [l] {0.5\textwidth}
\begin{center}
\epsfig{file=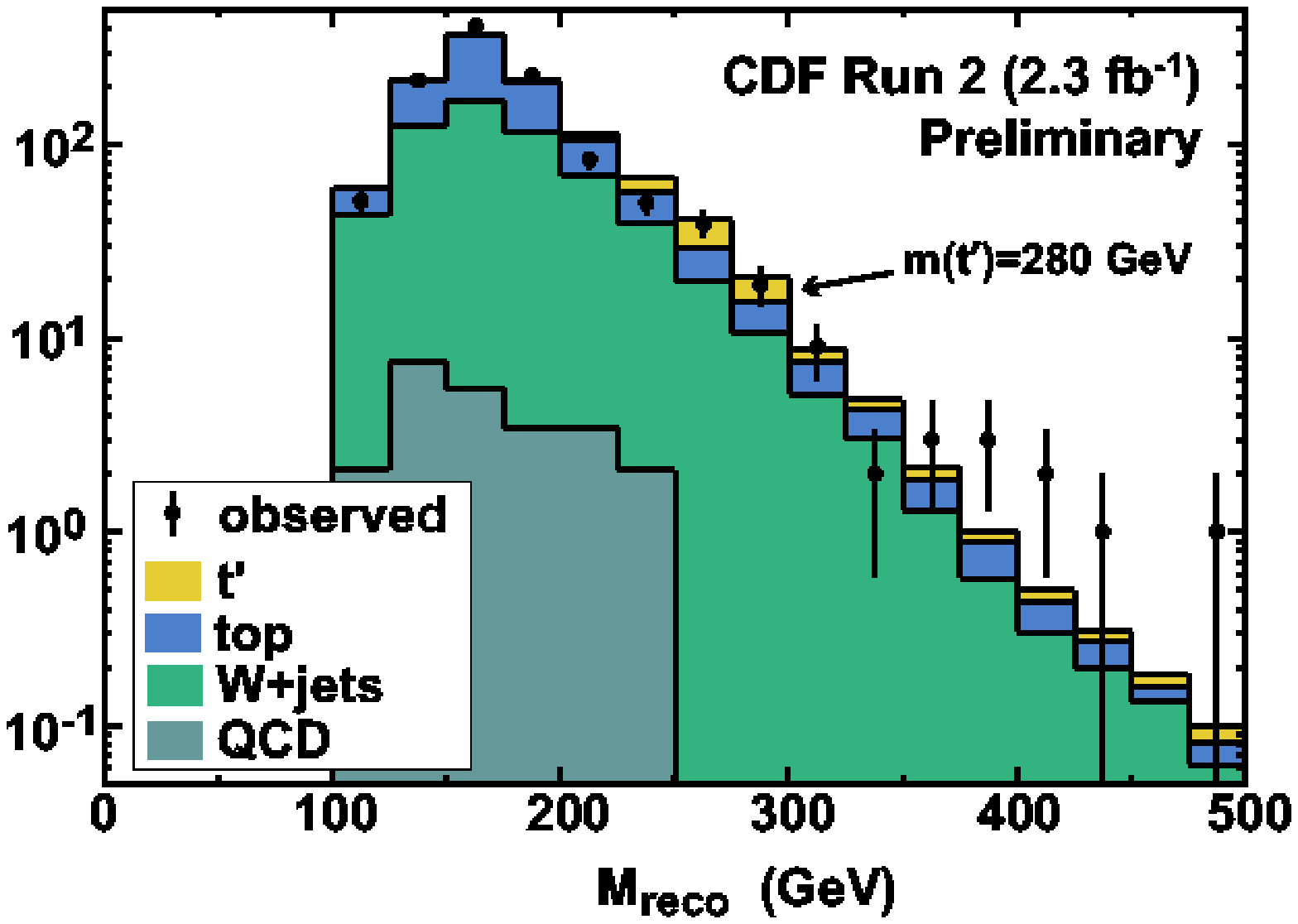, width=3in}
\caption{CDF Z $t^\prime$ mass spectrum.}
\label{fig:cdf_tprime}
\end{center}
\end{minipage}
\hspace{0.1cm}
\begin{minipage} [l] {0.5\textwidth}
\begin{center}
\epsfig{file=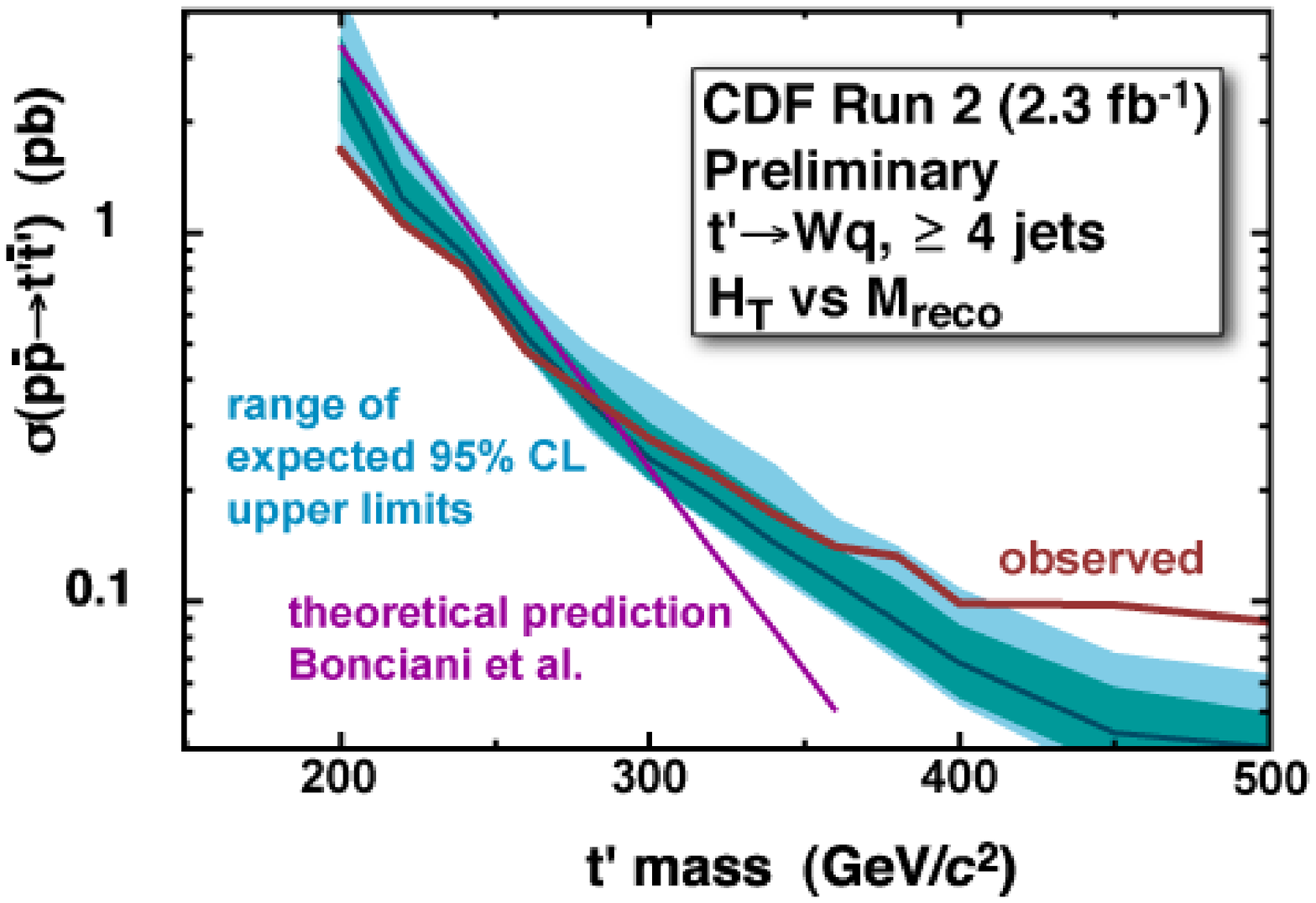, width=3in}
\caption{CDF Z $t^\prime$  exclusion limit.}
\label{fig:cdf_tprime_b}
\end{center}
\end{minipage}
\end{figure}

Part of the emerging `top quark factory' output are searches for new particles coupling predominantly to top quarks, like members of the fourth generation, \index{Fourth Generation} or new scalars leading to `maximal flavour violation' (see below). CDF has searched for a $t^\prime$ 
\index{$t^\prime$ }in $2.3\,fb^{-1}$ of data \cite{cdf:tprime}, with the expected outcome, and is excluding  $t^\prime$ masses below 284\,GeV (Fig.~\ref{fig:cdf_tprime}).

\subsection{Maximal Flavour Violation \index{Maximal Flavour Violation}}
New scalars decaying to like sign top quark pairs \index{like sign top quark pairs} are being excluded in the mass range from 180 to 300\,GeV for couplings to the top quark above 0.79 (180\,GeV) to 1.32 (300\,GeV) in a new CDF analysis \cite{cdf:maxflav}. 

\section{Global/Model-Independent Searches \index{model-independent searches} \index{global searches}}
\subsection{Excited Electrons \& Quarks}
CDF \cite{cdf:dijet} has searched for bumps \index{dijet resonances} in the dijet mass spectrum using $1.13\,fb^{-1}$ of data, and excludes excited quarks \index{excited quarks} 
with masses between 260 and 870\,GeV. \Dzero\  published a search \cite{dzero:estar} for exited electrons \index{excited electrons} decaying to $e \gamma$; based on data from $1\,fb^{-1}$ they exclude $e^*$ up to 756\,GeV, for a compositeness scale of 1 TeV.

\begin{figure}[htb]
\begin{minipage} [l] {0.5\textwidth}
\begin{center}
\epsfig{file=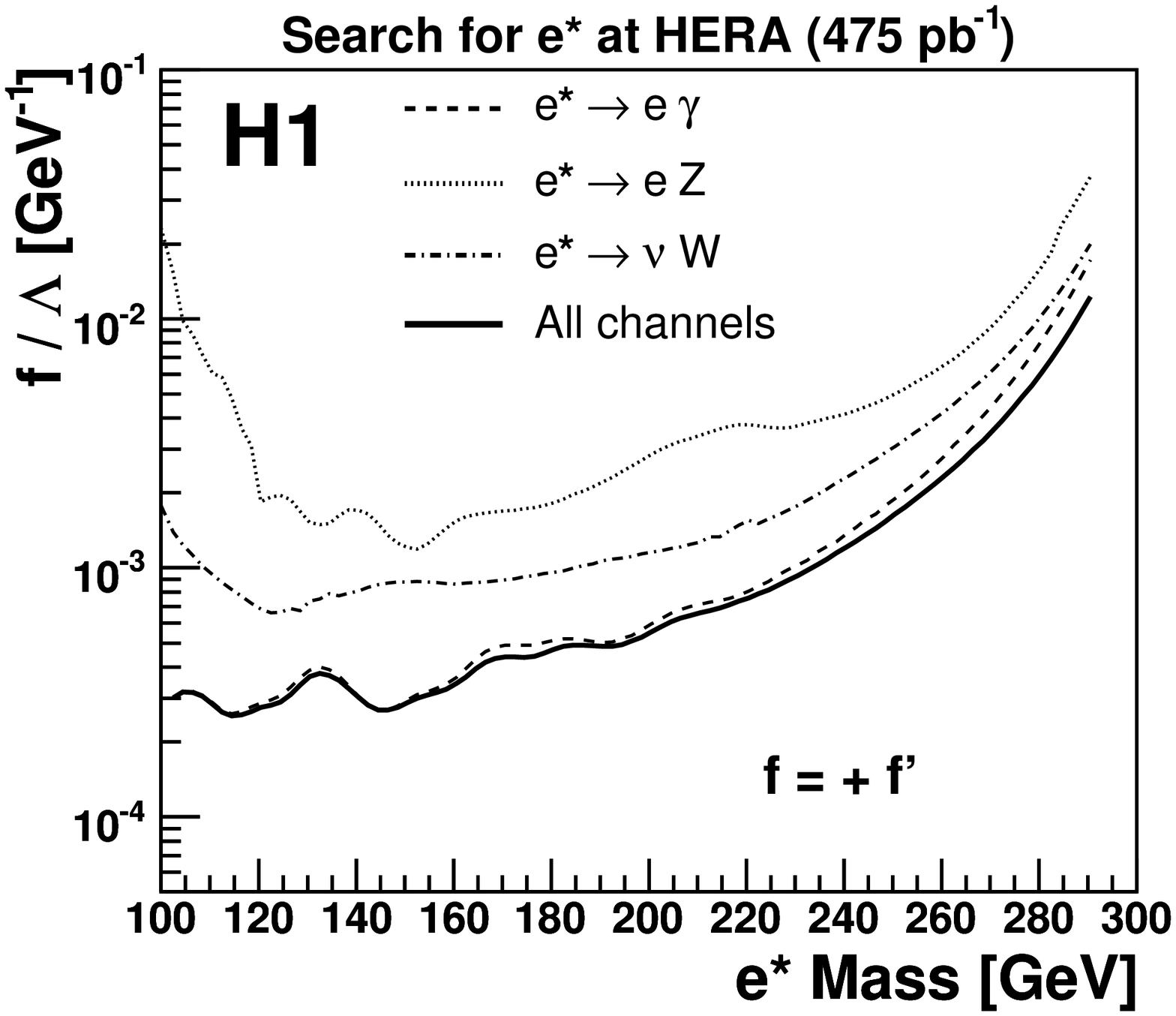, width=3in}
\caption{H1 $e^*$ exclusion region: breakdown by channel.}
\label{fig:H1_estar}
\end{center}
\end{minipage}
\hspace{0.1cm}
\begin{minipage} [l] {0.5\textwidth}
\begin{center}
\epsfig{file=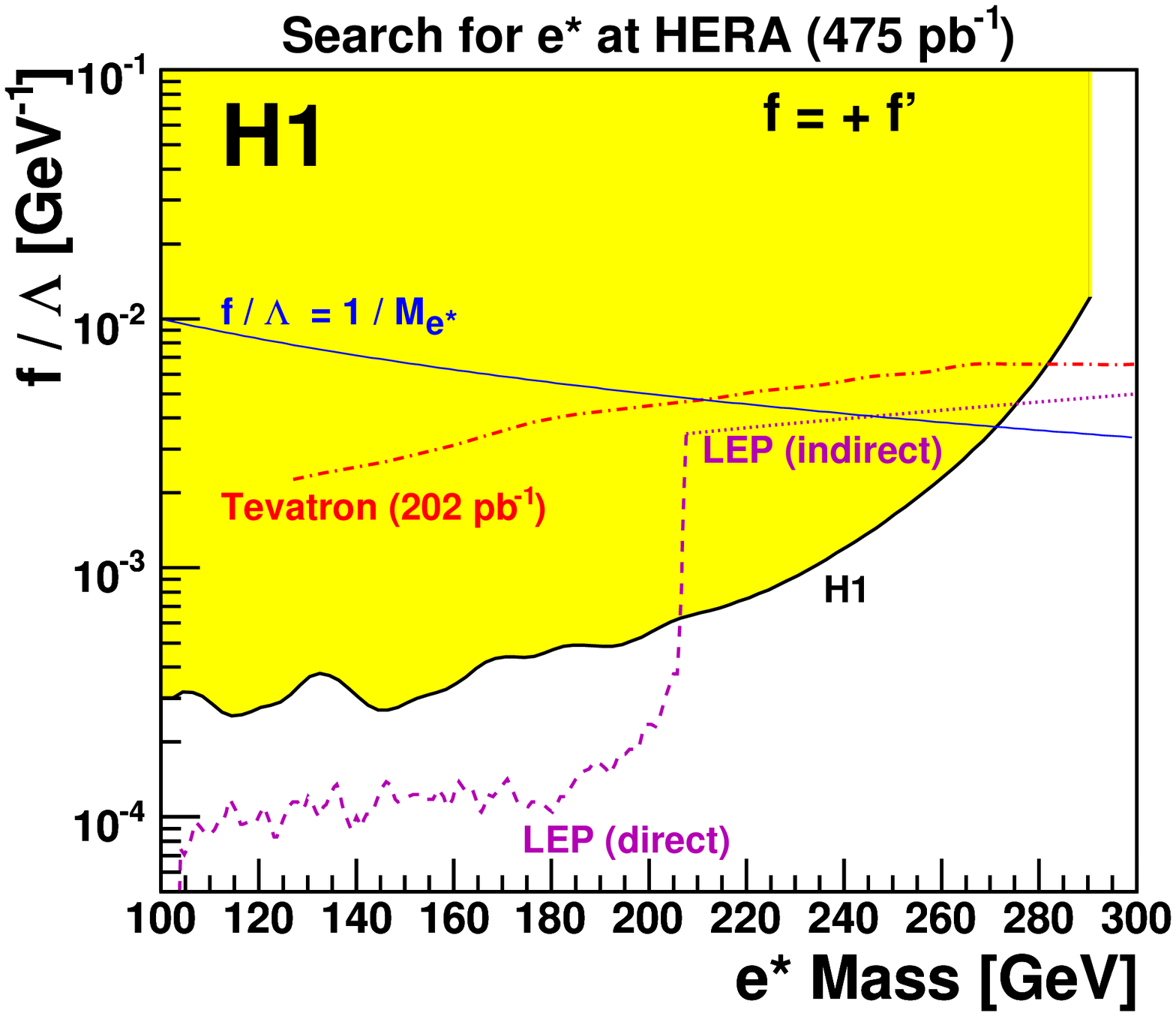, width=3in}
\caption{H1 $e^*$ exclusion region: comparison to other limits.}
\label{fig:H1_estar_b}
\end{center}
\end{minipage}

\end{figure}

At Hera,  lepton compositeness \index{lepton compositeness} can be studied via gauge mediated or contact interaction production of excited electrons and neutrinos. In 
\cite{H1:estar} the H1 collaboration describes a search for excited electrons decaying into $e \gamma$, $e Z$ or $\nu W$, using the full statistics accumulated at Hera. No deviations from the standard model are found, and a region is excluded in the plane of
the mass of the composite particle versus $f / \Lambda$, the ratio of the (SU(2)=U(1)) coupling strength to the compositeness scale  
(Fig.~\ref{fig:H1_estar}).

The companion paper \cite{H1:nustar} \index{excited neutrinos} sets limits resulting from a H1 search for excited neutrinos, based on $184\,pb^{-1}$ of $e^- p$ running at Hera.

 \subsection{Isolated and Multi-Leptons}
 After an initial report \cite{H1:isolep_excess} by H1 of a $3 \sigma$ excess in the production of isolated leptons, an update using the full Hera 
 statistics of $ 504\,pb^{-1}$ is now available from Zeus \cite{Zeus:isolep}. No significant excess is found. \index{Isolated leptons at HERA}
 
 Both H1 and Zeus have updated their high-$p_T$ multilepton studies to the full Hera statistics, and report no significant excess beyond standard model sources \cite{Hera:mullep}. \index{Multileptons at HERA}

\subsection{Long-lived Particles decaying to Diphotons or Dielectrons at the Tevatron}
\begin{figure}[htb]
\begin{minipage} [l] {0.5\textwidth}
\begin{center}
\epsfig{file=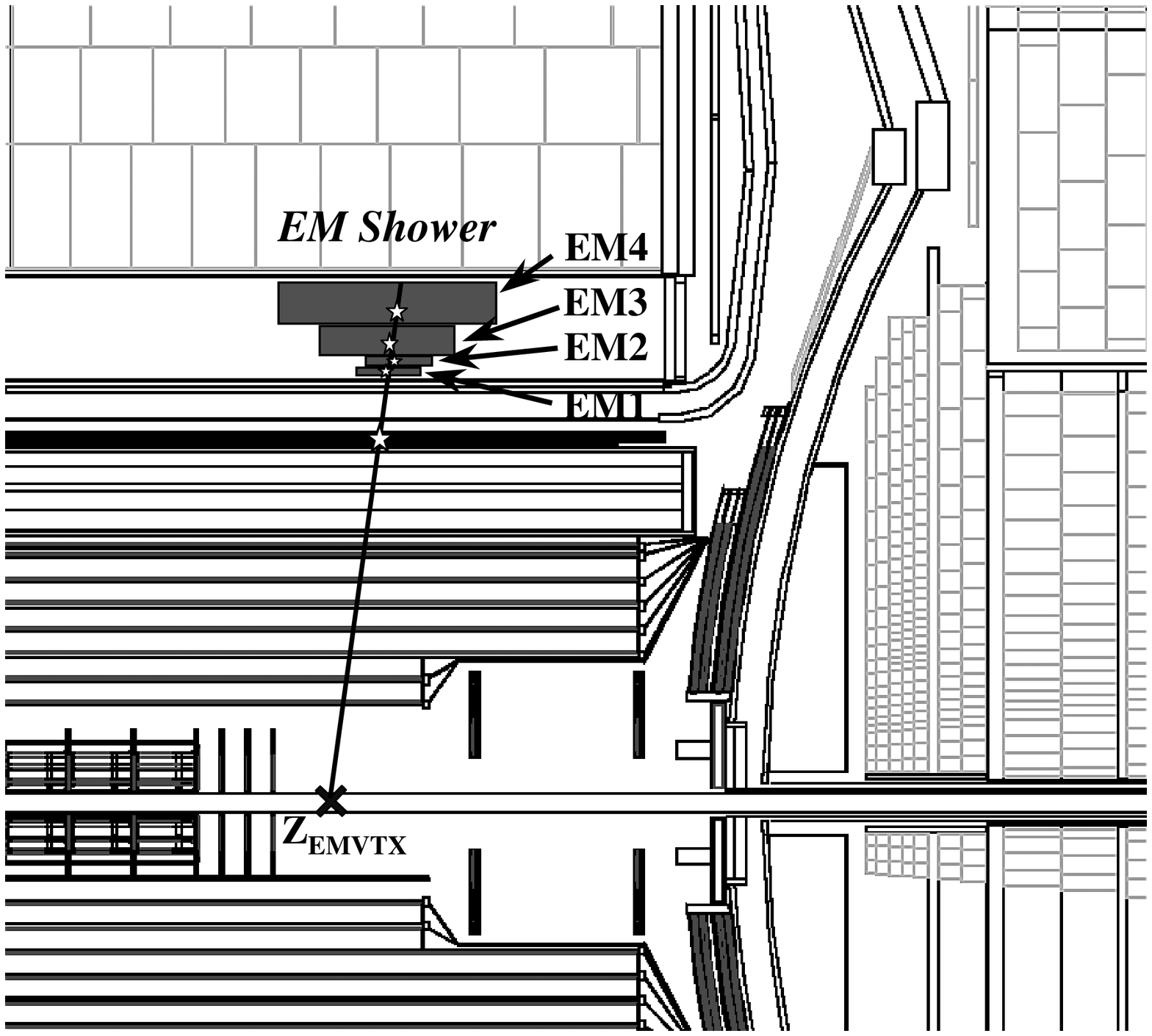,width=2.5in}
\caption{\Dzero\ EM pointing technique}
\label{fig:EM_pointing}
\end{center}
\end{minipage}
%\end{figure}
\hspace{0.1cm}
\begin{minipage} [l] {0.5\textwidth}
%\begin{figure}[htb]
\begin{center}
\epsfig{file=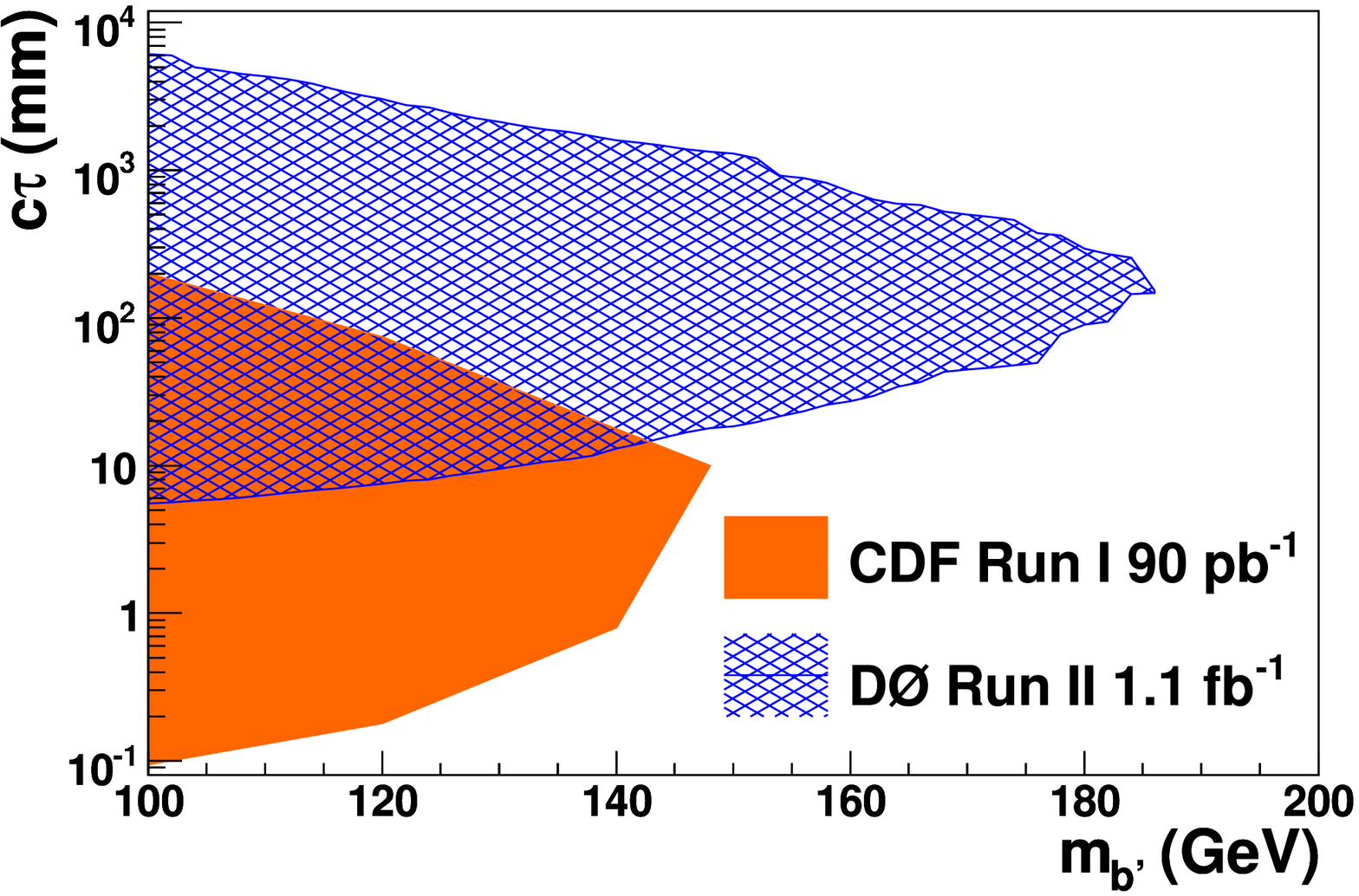,width=3in}
\caption{\Dzero\  exclusion contour for long-lived particles decaying to two photons or electrons.}
\label{fig:dzerolonglived}
\end{center}
\end{minipage}
\end{figure}

\Dzero\ has updated its search for long lived particles \index{long-lived particles} \index{Diphotons} \index{Dielectrons} 
yielding two electromagnetic clusters \cite{dzero:loli}. Using the EM pointing technique \index{EM pointing} \index{photon pointing} described above removes any dependence on tracking efficiency and thus extends the decay lengths covered significantly over a similar CDF analysis, despite the much smaller tracking volume (Fig.~\ref{fig:dzerolonglived}).

\subsection{Global Spectra Analysis and Bump Hunt at the Tevatron}
Initially conceived at \Dzero\  as Sherlock/Sleuth \index{Sherlock} \index{Sleuth} and used for beyond-the-Standard-Model searches at \Dzero\ and H1, model-independent searches for EW-scale new phenomena have received a big boost by the continued development at CDF. In a recent update \cite{cdf:vista} CDF describes the search for deviations from the standard model using the VISTA/Sleuth \index{VISTA} global comparison machinery. Using VISTA, roughly 4 million events from 
$2\,fb^{-1}$ were partitioned into about 400 exclusive final states, and the populations of those states as well as about 20,000 kinematic distributions were compared to standard model predictions. About 5000 mass spectra were scoured for mass bumps. No significant deviations from the Standard Model were found, after initial deficiencies in the event modeling and the detector description (in particular the imperfect modeling of 3-jet final states) were accounted for.

Sleuth was then used to specifically examine the high $p_T$ tails of 87 kinematic distributions for any excess not explained by SM sources. 
The probability to observe deviations as large or larger than the ones seen due to purely SM sources was quantified as 8\%. Thus, no new physics is required yet.

\section{Summary}
A very vibrant and diverse program of searches for physics beyond the Standard Model has led to many updates in the last 12 months 
from the large data sets amassed by the Hera and Tevatron collaborations. The final results from Hera, based on integrated luminosities 
of 500/pb for each of H1 and Zeus, have begun to arrive. Tevatron updates  based on up to 3 events per fb have been shown. 
With both CDF and \Dzero\  aiming for total integrated luminosities of $8\,fb^{-1}$, a lot more data are expected. 
These ever increasing data samples are complemented by advances in analysis techniques, pushing the limits of discovery ever further.

\def\Discussion{
\setlength{\parskip}{0.3cm}\setlength{\parindent}{0.0cm}
     \bigskip\bigskip      {\Large {\bf Discussion}} \bigskip}
\def\speaker#1{{\bf #1:}\ }
\def\endDiscussion{}

%\Discussion

%\speaker{D. Giovanni (University of Seville)}  My analysis indicates that the
%recovery of the two gentlemen is due simply to their embrace of the masculine
%principle and has nothing to do with magnetism at all.  Could you comment on 
%this?

%\speaker{Reggiano} Professor Giovanni has discussed this hypothesis in several
%forums, but, I do not believe there is anything in print.  I understand that
%he is spending his time in other pursuits.

%\speaker{D. Anna (University of Seville)}  In fact, my colleague Giovanni 
%has expressed opposite opinions on this question at various times, depending
%on the audience.  All of these testosterone-based theories are, of course,
%nonsense.

%\endDiscussion
 
\end{document}